\documentclass{article} 
\usepackage{latexsym}

\begin{document}

{\bf\huge\centerline{{Goryachev-Chaplygin, Kovalevskaya, and }}}
{\bf\huge\centerline{{Brdi\v{c}ka-Eardley-Nappi-Witten pp-waves spacetimes}}}
{\bf\huge\centerline{{with higher rank St\"{a}ckel-Killing tensors}}}

\bigskip
\bigskip
\bigskip
\bigskip
\bigskip

{\it\centerline{\LARGE{G. W. Gibbons {\footnote{gwg1@damtp.cam.ac.uk}}}}}
\medskip
{\it\centerline{\large{D. A. M. T. P.}}}
\medskip
{\it\centerline{\large{Wilberforce Road}}}
{\it\centerline{\large{Cambridge, CB3 9ET, UK}}}

\bigskip
\noindent
{\centerline{and}}
\bigskip

{\it\centerline{\LARGE{C. Rugina {\footnote{cristina.rugina@cea.fr}}}}}
\medskip
{\it\centerline{\large{Physics Department}}}
{\it\centerline{\large{University of Bucharest}}}
\medskip
{\it\centerline{\large{Bucharest,Romania}}}

\bigskip
\bigskip

{\bf\centerline{\LARGE{ABSTRACT}}}

\bigskip
\noindent
Hidden symmetries of the Goryachev-Chaplygin and Kovalevskaya gyrostats spacetimes, as well as the Brdi\v{c}ka-Eardley-Nappi-Witten pp-waves are studied. We find out that these spacetimes possess higher rank St\"{a}ckel-Killing tensors and that in the case of the pp-wave spacetimes the symmetry group of the St\"{a}ckel-Killing tensors is the well-known Newton-Hooke group.

\newpage
\section{Introduction}

\bigskip
\noindent
Hidden symmetries have been studied intensely since it was discovered that the associated rank-2 Killing-Yano and St\"{a}ckel-Killing tensors help separate the Dirac equation and the Hamilton-Jacobi and Klein-Gordon equations in various backgrounds, from the Kerr metric to more recently the Kerr-NUT-AdS spacetime [1 and references therein]. Recently a number of cases have been identified of spacetimes that admit higher rank irreducible St\"{a}ckel-Killing tensors - namely the Eisenhart lifts[2] of integrable systems Goryachev-Chaplygin[3,4], Kovalevskaya tops[5,6] and a pp-wave supersymmetric spacetime[7].

\medskip
\noindent
The classical problem of how many integrals of a geodesic flow exist for a given metric was posed since Whittaker, who proved that such an integral is equivalent to existence of an integral polynomial in momenta. Finding the integrals is a hard problem, since the integrals are usually analytic only on $T^*(M)\backslash M$. Elsewhere [8] Cartan's prolongation-projection scheme is used to find integrals of motion that are quadratic in momenta. We are going to follow the approach discussed by van Holten [9] and Visinescu [10] for finding higher degree polynomial constants of motion- for motion of particles in external fields- the case discussed here is the Brdi\v{c}ka-Eardley-Nappi-Witten pp-wave[11, 14, 15]. For constants of motion to exist in the case of particles in external fields, the symmetries of the metric and those of the external fields have to match.

\medskip
\noindent
The Eisenhart lifts studied recently[7] have well-known integrable rigid-body system dynamics and their quantum analogs are constructed based on well-known group behavior- as for instance is the case with the Goryachev-Chaplygin top and the E(3) group that describes its quantum behaviour. Another well-known Liouville integrable system is the Kovalevskaya top- studied in its classical aspects beginning from the end of the nineteenth century.

\medskip
\noindent
In this paper we look at other integrable systems: the Goryachev-Chaplygin[12] and Kovalevskaya gyrostats [13] - (GCG) and (KG) respectively, their quantum counterparts and the corresponding lifted spacetimes. In close resemblance to their top cousins, they exhibit hidden symmetries governed by higher rank St\"{a}ckel-Killing tensors, which are the lifted constants of motion in the classical case. The gyrostat terms added to the (classical) quadratic Hamiltonian do not modify the Poisson brackets algebra from that of the top case, and again the hamiltonian can be considered as generating the geodesic flow of a (pseudo)-Riemannian metric. The Schouten-Nijenhuis algebra of the Killing tensors in the lifted spacetime will be the same as the Poisson algebra of the constants of the motion for the original dynamical system. The degree of integrability of the dynamical systems is unchanged by the dynamical lift.

\medskip
\noindent
For (GCG) the separation of variables is based on the connection between the degenerated representation of the E(3) group and the special representation of the SO(3,2) group. The SO(3,2) has a semisimple subgroup SO(2,2), which is a dynamical group for (GCG). For (KG) the integrals of motion were constructed on o(4), e(3) and o(3,1) Lie algebras[13]. Here we work with the case where the Lie algebra is e(3).

\medskip
\noindent
The Brdi\v{c}ka-Eardley-Nappi-Witten pp-wave results from an Eisenhart lift of a general class of pp-wave solutions to the Einstein-Maxwell equations, requiring that the electromagnetic tensor is covariantly constant. Later the same spacetime was constructed by Nappi and Witten as the target space of a Wess-Zumino-Witten conformal field theory. This spacetime is completely homogeneous and is a bi-invariant metric on the Cangemi-Jackiw group, as it has been shown by Gibbons and Pope[11]. It is shown that this spacetime admits four rank-3 St\"{a}ckel-Killing tensors and that the spacetime of St\"{a}ckel-Killing tensors is symmetrical under Newton-Hooke group.

\medskip
\noindent
The paper is structured as follows: sections 2 and 3 are devoted to the (GCG) - the classical and quantum cases, sections 4 and 5 to the (KG)- the classical and quantum cases and section 6 to the pp-wave spacetime. We end with conclusions.

\section{The classical Goryachev-Chaplygin gyrostat} 

\noindent
We consider the generalization of the integrable classical Goryachev-Chaplygin top, the (GCG). The hamiltonian differs from that of the corresponding top by linear terms in the body frame, depending additionally on $\lambda$, a constant arbitrary vector, the gyrostatic moment:

\begin{equation}
H_{GCG} = \frac{1}{2} (m^2_1 + m^2_2 + 4 m_3^2) + \alpha^2 sin\theta sin \psi -2 m_3 \lambda + 2\lambda^2
\end{equation}

\noindent
and the corresponding constant of motion -

\begin{equation}
K_{GCG} = (m_3-2\lambda) (m_1^2 +m_2^2+ \frac{1}{4}) -\alpha^2 m_2 cos \theta
\end{equation}

\noindent
and $m_1, m_2, m_3$ are as follows:

\begin{equation}
m_1 = - sin\psi p_{\theta} - cos \psi cot \theta p_{\psi}, \hspace{0.25in}
m_2 = cos\psi p_{\theta} - sin\psi cot\theta p_{\psi}, \hspace{0.25in} m_3 = p_{\psi}.
\end{equation}

\noindent
This generates the geodesic flow of the 4-dimensional metric of this spacetime with Killing vector fields $k=\partial_t$ and $l= \partial_s$, where l is lightlike and covariantly constant:

\begin{equation}
g = (-2 \alpha^2 sin \theta sin \psi -8\lambda^2) dt^2 +2dt ds + d\theta^2 + 
\frac{d\psi^2}{cot^2\theta +4} - \frac{1}{8 \lambda}d\psi dt.
\end{equation}

\noindent
The lifted classical GCG hamiltonian is -

\begin{equation}
\mathcal{H}_{GCG} = (m_1^2 + m_2^2 + 4 m_3^2) + 2 \alpha^2 sin \theta sin \psi p^2_s + 2 p_s p_t - 2 m_3 \lambda p_s + 2 \lambda^2 p_s^2
\end{equation}

\noindent
and the lifted (in 4 dimensions) classical St\"{a}ckel-Killing tensor is-

\begin{equation}
\mathcal{K}_{CGC} = m_3 (m^2_1 + m^2_2) - \alpha^2 p^2_s m_2 cos\theta - 2 \lambda(m^2_1 + m^2_2)p_s - \frac{\lambda}{2} p_s^3 + \frac{1}{4}m_3p_s^2
\end{equation}

\noindent
with components

\begin{eqnarray}
K^{\theta \theta \psi} = \frac{1}{3}, \hspace{.25in}
K^{\theta ss} =- \frac{\alpha^2}{3} cos \psi cos \theta, \hspace{.25in}
K^{\psi\psi\psi} = cot^2 \theta, \\
K^{\psi ss} = \frac{\alpha^2}{3} \frac{cos^2\theta sin\psi}{sin \theta} +\frac{1}{4}, \hspace{.25in} K^{\theta \theta s} = -2\lambda, \hspace{0.25in}
K^{\psi \psi s} = -2 \lambda cot^2 \theta.
\end{eqnarray}

\noindent
One may verify that this tensor satisfies the Killing equation: $\nabla^{(a} K^{bcd)} = 0$. Among the properties of this spacetime: it isn't Ricci flat and the Ricci scalar does not vanish; consequently, it does not admit a Killing spinor according to the classification of manifolds that admit Killing spinors.

\noindent
\section{ The quantum mechanical Goryachev-Chaplygin gyrostat} 

\noindent
The quantum mechanical top and gyrostat has been studied by Komarov and collaborators in a number of papers (see ref. [12] and refrences therein)
The quantum mechanical hamiltonian reads (cf [12]):

\begin{equation}
\hat{H}_{GCG} = \frac{1}{2} (J^2_1 + J^2_2 + 4 J^2_3) -  \alpha^2 x_2 + 8 \lambda J_3 + 4 \lambda^2. 
\end{equation}

\noindent
The $\hat{K}_{GCG}$ operator for the Goryachev-Chaplygin gyrostat reads:

\begin{equation}
\hat{K}_{GCG} = J_3 (J^2_1 + J^2_2) -\frac{1}{4} J_3 - \frac{1}{2} \alpha^2 (J_2 x_3 + x_3 J_2) - 2 \lambda (J^2_1 + J^2_2) + \frac{\lambda}{2}.
\end{equation}

\noindent
Applying the Eisenhart lift to the above hamiltonian gives the following one in 5 dimensions with the following metric (with notations in [7]):

\begin{equation}
g_{GCG} = 2 ds dt + 2(4\lambda^2 + \alpha^2 x_2)dt^2 - 16 \lambda ds \sigma^3 + (\sigma^1)^2 + (\sigma^2)^2 + \frac{1}{4}(\sigma^3)^2,
\end{equation}

\begin{equation}
\mathcal{H}_{GCG} = (J^2_1 + J^2_2 + 4 J^2_3) - 16 \lambda J_3 \partial_s + 2 (4 \lambda^2 + \alpha^2 x_2) \partial_s^2 + 2 \partial_s \partial_t.
\end{equation}

\noindent
And the lifted $\mathcal{K}_{GCG}$ operator that commutes with this hamiltonian reads:

\begin{equation}
\mathcal{K}_{GCG} =  J_3(J^2_1 +J^2_2) - \frac{1}{4} J_3 - \frac{\alpha^2}{2} \{x_3, J_2\}\partial_s^2 -2 \lambda (J^2_1 + J^2_2)\partial_s + \frac{\lambda}{2}\partial_s^3.
\end{equation}

\noindent
Moreover, this operator can be written as:

\begin{equation}
\mathcal{K}_{GCG} = K^{abc}_{GCG} \nabla_a \nabla_b \nabla_c + \frac{3}{2} (\nabla_a K^{abb}_{GCG}) \nabla_b \nabla_b -\frac{1}{2} K_{(GCG) a}^{ab} \nabla_b
\end{equation}

\noindent
and $K_{GCG}$ is a symmetric rank 3 tensor with the following non-zero components:

\begin{eqnarray}
K^{113} = K^{223} = \frac{2}{3}, \hspace{.25in}
K^{11s} = K^{22s} = -\frac{4 \lambda}{3}, \\
K^{2ss} = -\frac{2 \alpha^2}{3} x_3, \hspace{.25in}
K^{sss} = \frac{\lambda}{3}.
\end{eqnarray}

\noindent
Introducing the basis: 

\begin{equation}
L_s = \partial_s, \hspace{0.25in} L_t = \partial_t, \hspace{0.25in} L_1 = J_1, \hspace{0.25in} L_2= J_2, \hspace{0.25in} L_3 =J_3-\lambda,
\end{equation}

\noindent
one finds that the tensor satisfies $\nabla^{(a} K_{GCG}^{bcd)} = -\alpha^2 L^{(a}_s L^b_s (\partial_\phi)^c L^{d)}_1$, which if $p_\phi$ vanishes, gives the Killing equation.

\noindent
For the naive quantization in 4 dimensions of the lifted classical case, we get:

\begin{equation}
\hat{H}_{GCG} = \frac{1}{2} (j^2_1 + j^2_2 + 4 j^2_3) -  \alpha^2 x_2 + 8 \lambda j_3 + 4 \lambda^2. 
\end{equation}

\noindent
The $\hat{K}_{GCG}$ operator for the Goryachev-Chaplygin gyrostat reads:

\begin{equation}
\hat{K}_{GCG} = j_3 (j^2_1 + j^2_2) -\frac{1}{4} j_3 - \frac{1}{2} \alpha^2 (j_2 x_3 + x_3 j_2) - 2 \lambda (j^2_1 + j^2_2) + \frac{\lambda}{2},
\end{equation}

\noindent
where we have defined the operators:

\begin{equation}
j_1 = - sin\psi \partial_\theta - cos\psi cos\theta \partial_\psi, \hspace{0.25in} j_2 = cos\psi \partial_\theta - sin\psi cot \theta \partial_\psi,
\hspace{0.25in} j_3= \partial_\psi
\end{equation}

\noindent
By lifting the hamiltonian and K operators, one finds:

\begin{equation}
\mathcal{H}_{GCG} = (j^2_1 + j^2_2 + 4 j^2_3) - 16 \lambda j_3 \partial_s + 2 (4 \lambda^2 + \alpha^2 x_2) \partial_s^2 + 2 \partial_s \partial_t
\end{equation}

\noindent
and

\begin{equation}
\mathcal{K}_{GCG} =  j_3(j^2_1 + j^2_2) - \frac{1}{4} j_3 - \frac{\alpha^2}{2} \{x_3, j_2\}\partial_s^2 -2 \lambda (j^2_1 + j^2_2)\partial_s + \frac{\lambda}{2}\partial_s^3.
\end{equation}

\noindent
One can verify that $[\mathcal{H}_{CGC}, \mathcal{K}_{CGC}]=0$. Also it turns out that for the GCG spacetime with the metric given by (3),

\begin{equation}
\Box = \mathcal{H} -\frac{3 cot \theta}{4+ cot^2 \theta}\partial_\theta - \frac{3}{4\lambda} \partial_\psi.
\end{equation}

\bigskip
\noindent
\section {The classical Kovalevskaya gyrostat} 

\bigskip
\noindent
In late nineteenth century Madame Kovalevskaya found and integrated a new integrable case of rotation of a heavy rigid body around a fixed point. Nowadays we think of this as an integrable system on the orbits of the Euclidean Lie algebra e(3) with a constant hamiltonian and a quartic in momenta integral of motion. We extend the study of this system, dubbed Kovalevskaya top to a generalization (fluid-filled cavities rigid body- where the Coriolis force becomes important) - the Kovalevskaya gyrostat (KG).

\medskip
\noindent
In this case the hamiltonian writes:

\begin{equation}
H = \frac{1}{2}(m_1^2 + m_2^2 + 2 (m_3 - \lambda)^2)+ \alpha^2 x_1
\end{equation}

\noindent
and commutes with the constant of motion

\begin{equation}
K = (m_1^2 + m_2^2)^2 + 4 \alpha^4 (x_1^2 + x_2^2) - (4 \alpha^2 x_1 + 8 \lambda^2)(m_1^2 -m_2^2) -8\alpha^2 x_2 m_1 m_2,
\end{equation}

\noindent
where $x_1 = sin \theta cos \psi$ and $ x_2 = sin \theta sin \psi$. The above mentioned hamiltonian lifts to:

\begin{equation}
\mathcal{H}_{GCG} = (m_1^2 + m_2^2 + 2 m_3^2) + 2 \alpha^2 x_1 p^2_s + 2 p_s p_t - 4 m_3 \lambda p_s + 2 \lambda^2 p_s^2,
\end{equation}

\noindent
which generates geodesic flow of the following Lorentzian 4-metric-

\begin{equation}
g = (-2 \alpha^2 sin \theta sin \psi -16 \lambda^2) dt^2 +2dt ds + d\theta^2 + 
\frac{d\psi^2}{cot^2\theta + 2} - \frac{1}{8 \lambda}d\psi dt.
\end{equation}

\noindent
This admits the rank - 4 irreducible tensor K (the lifted constant of motion):

\begin{equation}
K = (m_1^2 + m_2^2)^2 + 4 \alpha^4 (x_1^2 + x_2^2) p_s^4 -[( 4 \alpha^2 x_1 + 8\lambda^2)(m_1^2 -m_2^2)  + 8 \alpha^2 x_2 m_1 m_2] p_s^2,
\end{equation}

\noindent
with components-

\begin{eqnarray}
K^{\theta \theta \theta \theta} = 1, \hspace{.25in} 
K^{\theta \theta \psi \psi} = \frac{1}{3} cot^2 \theta, \hspace{.25in}
K^{ss \theta \theta} = \frac{2}{3} \alpha^2 sin \theta cos \psi - 8 \lambda^2(sin^2 \psi -cos^2 \psi), \\
K^{ss \theta \psi} = - \frac{2}{3} \alpha^2 cos\theta sin\psi, \hspace{.25in} K^{ssss} = 4 \alpha^4 sin^2 \theta, \\
K^{ss\psi \psi} = -\frac{2}{3} \alpha^2 cos\psi cos \theta cot \theta - 8\lambda^2(cos^2\psi -sin^2 \psi)cot^2 \theta. 
\end{eqnarray}

\noindent
We find in accordance with [7] that the properties of the (KG) spacetime are very similar to those of the Kovalevskaya top spacetime and to those of (GCG), namely it is not Ricci flat and does not admit a Killing spinor. In this 4-dimensional context also we note that quantum mechanically  $\Box \neq \mathcal{H}$, where $\mathcal{H}$ is obtained by the 'naive' quantization procedure used in section 3.

\bigskip
\noindent
\section{ The quantum mechanical Kovalevskaya gyrostat} 

\bigskip
\noindent
The quantum mechanical Kovalevskaya top (KT) was studied in [13] and references therein. The integrals of motion for the KT form the basis for the hydrogen atom. Here we continue to explore higher rank St\"{a}ckel-Killing tensors for the (KT) generalization.

\medskip
\noindent
We consider a 5D spacetime for quantization with a metric given by:

\begin{equation}
g_{KG}= 2 (2\lambda^2 + \alpha^2 x_1) dt^2 + 2 ds dt - 8 \lambda ds \sigma^3 + (\sigma^1)^2 + (\sigma^2)^2 + \frac{1}{2} (\sigma^3)^2
\end{equation}

\noindent
and a hamiltonian given by [13]-

\begin{equation}
\Box_{KG} = H_{KG} = \frac{1}{2} (J^2_1 + J^2_2 + 2(J_3 - \lambda)^2) + \alpha^2 x_1
\end{equation}

\noindent
and a constant of motion operator that commutes with the hamiltonian

\begin{equation}
K_{KG} = \frac{1}{2}\{K_{+}, K_{-}\} - \lambda [K_{+}, K_{-}] +(2 - 8 \lambda^2)\{J_{+}, J_{-}\},
\end{equation}

\noindent
where $J_{\pm} = J_1 \pm i J_2$, $K_{\pm} = J^2_{\pm} - 2 \alpha^2 x_{\pm} \partial ^2_s$ and $x_{\pm} = x_1 \pm i x_2$. In the basis

\begin{equation}
L_s = \partial_s, \hspace{0.25in} L_t = \partial_t, \hspace{0.25in} L_1 = J_1, \hspace{0.25in} L_2= J_2, \hspace{0.25in} L_3 =J_3-\lambda,
\end{equation}

\noindent
the five-dimensional rank-4 irreducible Killing tensor $K_{KG}$ is-

\begin{eqnarray}
\mathcal{K}_{KG} = K^{abcd}_{KG} \nabla_a \nabla_b \nabla_c \nabla_d + 2 (\nabla_a K^{abcd}_{KG}) \nabla_b \nabla_c \nabla_d + 3(\nabla_a \nabla_b K^{abcd}_{KG})\nabla_c \nabla_d \\
 - 2 {K^{abc}_{KG}}_c \nabla_a \nabla_b - \frac{3}{4} {K^{ab}_{KG}}_{ab} L^c_3 L^d_3 \nabla_c \nabla_d
\end{eqnarray}

\noindent
and in components it is-

\begin{eqnarray}
K^{ssss}_{KG} = 4 \alpha^4 (x_1^2 + x_2^2), \hspace{0.25in} K^{ss11}_{KG} = -K^{ss22}_{KG} = -2 \alpha^2 x_1/3 (1 - \lambda), \\
K^{ss12}_{KG} = -2 \alpha^2 x_1/3 (1 - \lambda), \hspace{0.25in}
K^{1111}_{KG} = 3 K^{1122}_{KG} = K^{2222}_{KG} = 1 - 8 \lambda^2.
\end{eqnarray}

\medskip
\noindent
As it was determined before[13], there is a correlation between the (GCG) and (KG) cases and their constants of motion. Also note that the Lax pair with spectral parameter accounts for the gyrostatic term in the hamiltonian. 

\bigskip
\noindent
\section{ The Brdi\v{c}ka-Eardley-Nappi-Witten pp-waves }

\bigskip
\noindent
We continue to investigate spacetimes that admit higher ranking St\"{a}ckel-Killing tensors, such as the Brdi\v{c}ka-Eardley-Nappi-Witten pp-waves spacetime, which is an exact solution of the Einstein-Maxwell equations. We firstly take a look at the classical Hamiltonian of the form:

\begin{equation}
\mathcal{H} = \frac{1}{2m} g^{IJ} (\pi_I - e A_I) (\pi_J - e A_J) + V(x)
\end{equation}

\noindent
with V given by-

\begin{equation}
V = \frac{1}{2} m \omega^2 \sum_a r^2_a + \sum_{a,b} V(r_a - r_b),
\end{equation}

\noindent
which admits a number of Killing vectors according to [11] and four rank three St\"{a}ckel-Killing tensors determined as solutions to the differential equation - following the scheme in [9,10]. They describe hidden symmetries of the spacetime. One of them has the following components - with notations in [11] for $\Omega$ and $\nu$:

\begin{eqnarray}
K^{x_a x_a x_a} = \frac{2}{m} \frac{cos(\Omega + \nu)t}{\frac{\partial V}{\partial x_a}}, \hspace{0.25in} K^{y_a y_a y_a} = \frac{2}{m} \frac{sin(\Omega +\nu)t}{\frac{\partial V}{\partial y_a}}, \\
K^{x_a y_a x_a} = \frac{2}{m} \frac{sin(\Omega+\nu)t}{\frac{\partial V}{\partial x_a}}, \hspace{0.25in} K^{y_a x_a y_a} = \frac{2}{m} \frac{cos(\Omega + \nu)t}{\frac{\partial V}{\partial y_a}}.
\end{eqnarray}

\noindent
We lift the hamiltonian to:

\begin{equation}
\tilde{\mathcal{H}} = g^{IJ} (\pi_I - \frac{e}{m} A_I \pi_s)(\pi_J -\frac{e}{m} A_J \pi_s) + 2\pi_s\pi_t +\frac{2}{m} V \pi^2_s
\end{equation}

\noindent
with the higher dimensional metric-

\begin{equation}
ds^2 = g_{IJ} dx^I dx^J +\frac{2e}{m} A_I dx^I dt + 2dtdv -\frac{2}{m} V dt^2, 
\end{equation}

\noindent
which becomes in the center of mass frame restricted to z=0-

\begin{equation}
ds^2 = N[dx^2 + dy^2 + (\Omega^2(x^2 + y^2) + \omega^2 z^2)dt^2] +2dtds.
\end{equation}

\noindent
This metric is a particular case of the general pp-wave solution to the Einstein -Maxwell equation that is defined by two arbitrary holomorphic functions. With notations in[11] this metric is:

\begin{equation}
ds^2 = du(dv + H(u,\zeta, \bar{\zeta}) du) +d\zeta d\bar{\zeta}
\end{equation}

\noindent
and with a coordinate transformation given in [11]- the pp-wave Nappi-Witten metric becomes-

\begin{equation}
ds^2 = 2 du'dv' + b du'^2 + da^i da^j \delta_{ij} + \epsilon_{ij}da_i du'.
\end{equation}.

\noindent
The lifted rank-3 St\"{a}ckel-Killing tensor becomes - the components written above stay the same and the new non-null components are:

\begin{eqnarray}
K^{x_a s x_a} = \frac{2N \Omega}{m}[x_a sin(\Omega +\nu)t + y_a cos(\Omega +\nu)t] \frac{1}{\frac{\partial V}{\partial x_a}}, \\
K^{y_a s y_a} = \frac{2N \Omega}{m}[x_a sin(\Omega +\nu)t + y_a cos(\Omega +\nu)t] \frac{1}{\frac{\partial V}{\partial y_a}}.
\end{eqnarray}

\noindent
Note that in [11] and the above metrics t is u, s is v and $\zeta= x+iy$ and so the above components can be written as- in the center of mass metric:

\begin{eqnarray}
K^{x s x} = \frac{2N \Omega}{m}[\frac{(\zeta+\bar{\zeta})}{2} sin(\Omega +\nu)u + \frac{(\zeta-\bar{\zeta})}{2i} cos(\Omega +\nu)u] \frac{1}{\frac{\partial V}{\partial \zeta}}, \\
K^{y s y} = \frac{2N \Omega}{m}[\frac{(\zeta+\bar{\zeta})}{2} sin(\Omega +\nu)u + \frac{(\zeta-\bar{\zeta})}{2i} cos(\Omega +\nu)u] \frac{1}{\frac{-i \partial V}{\partial \zeta}}.
\end{eqnarray}

\noindent
In the Nappi-Witten pp-wave coordinates the components of the 4 rank-3 St\"{a}ckel-Killing tensors are:

\begin{eqnarray}
K^{x x x}_1 = i K^{y x y}_1= \frac{2}{m} \frac{cos(\Omega + \nu)\frac{u'}{\sqrt{8} C}}{\frac{\partial V}{\partial \zeta}}, \hspace{0.25in} K^{y y y}_1 = i K^{x y x}_1 = \frac{2}{m} \frac{sin(\Omega +\nu)\frac{u'}{\sqrt{8} C}}{\frac{-i\partial V}{\partial \zeta}}, \\
K^{x s x}_1 = i K^{y s y}_1= \frac{2N \Omega}{m}[\frac{(\zeta+\bar{\zeta})}{2} sin(\Omega +\nu)\frac{u'}{\sqrt{8} C} + \frac{(\zeta-\bar{\zeta})}{2i} cos(\Omega +\nu)\frac{u'}{\sqrt{8} C}] \frac{1}{\frac{\partial V}{\partial \zeta}},
\end{eqnarray}

\begin{eqnarray}
K^{x x x}_2 = i K^{y x y}_2 \frac{2}{m} \frac{sin(\Omega + \nu)\frac{u'}{\sqrt{8} C}}{\frac{\partial V}{\partial \zeta}}, \hspace{0.25in} K^{y y y}_2 = i K^{x y x}_2 = \frac{2}{m} \frac{cos(\Omega +\nu)\frac{u'}{\sqrt{8} C}}{\frac{-i\partial V}{\partial \zeta}}, \\
K^{x s x}_2 = i K^{y s y}_2= \frac{2N \Omega}{m}[-\frac{(\zeta+\bar{\zeta})}{2} cos(\Omega +\nu)\frac{u'}{\sqrt{8} C} + \frac{(\zeta-\bar{\zeta})}{2i} sin(\Omega +\nu)\frac{u'}{\sqrt{8} C}] \frac{1}{\frac{\partial V}{\partial \zeta}}, 
\end{eqnarray}

\begin{eqnarray}
K^{x x x}_3 = i K^{y x y}_3= \frac{2}{m} \frac{cos(\Omega - \nu)\frac{u'}{\sqrt{8} C}}{\frac{\partial V}{\partial \zeta}} ,\hspace{0.25in} K^{y y y}_3 = i K^{x y x}_3 = \frac{2}{m} \frac{sin(\Omega -\nu)\frac{u'}{\sqrt{8} C}}{\frac{-i\partial V}{\partial \zeta}}, \\
K^{x s x}_3 = i K^{y s y}_3= \frac{2N \Omega}{m}[\frac{(\zeta+\bar{\zeta})}{2} sin(\Omega -\nu)\frac{u'}{\sqrt{8} C} - \frac{(\zeta-\bar{\zeta})}{2i} cos(\Omega -\nu)\frac{u'}{\sqrt{8} C}] \frac{1}{\frac{\partial V}{\partial \zeta}}, 
\end{eqnarray}

\begin{eqnarray}
K^{x x x}_4 = i K^{y x y}_4 \frac{2}{m} \frac{-sin(\Omega - \nu)\frac{u'}{\sqrt{8} C}}{\frac{\partial V}{\partial \zeta}}, \hspace{0.25in} K^{y y y}_4 = i K^{x y x}_4 = \frac{2}{m} \frac{cos(\Omega -\nu)\frac{u'}{\sqrt{8} C}}{\frac{-i\partial V}{\partial \zeta}} ,\\
K^{x s x}_4 = i K^{y s y}_4= \frac{2N \Omega}{m}[\frac{(\zeta+\bar{\zeta})}{2} cos(\Omega -\nu)\frac{u'}{\sqrt{8} C} + \frac{(\zeta-\bar{\zeta})}{2i} sin(\Omega -\nu)\frac{u'}{\sqrt{8} C}] \frac{1}{\frac{\partial V}{\partial \zeta}}, 
\end{eqnarray}

\noindent
where $\zeta = (a_1 -ia_2) e^{\frac{iu'}{2}}$. The $\mathcal{K} = K^{a_1 a_2 a_3} p_{a_1} p_{a_2} p_{a_3}$ constants of motion derived from St\"{a}ckel-Killing tensors commute with the center of mass hamiltonian taken in the z=0 frame (planar case) and form a Poisson algebra accordingly:

\begin{equation}
\{K_i, K_j\} = K_k,
\end{equation}

\noindent
where i, j, k are circular permutations of 1 to 4. To be noted that the planar hamiltonian of the system in study admits (see [11]) a six-dimensional Newton-Hooke symmetry group, which is a Wigner -In\"{o}n\"{u} contraction of SO(2,2) and that in particular the Brdi\v{c}ka-Eardley-Nappi-Witten metric is bi-invariant on Cangemi-Jackiw group, so the isometry group of this metric is $(\mathcal{CJ} \bigotimes \mathcal{CJ})/N$. It turns out that the symmetry group of the vector space of all St\"{a}ckel-Killing tensors of the Brdi\v{c}ka-Eardley-Nappi-Witten spacetime is actually the Newton-Hooke group, in getting this we have generalized the considerations in [16,17]. In [17]-section 3- they define six transformations on the vector space of all St\"{a}ckel -Killing tensors defined on a Minkowski metric. In our case the Minkowski metric is replaced by the (BENW) pp-wave metric and so the underlying group becomes a continuous one-parameter family of deformations and after a careful analysis it turns out to be the Wigner-In\"{o}n\"{u} contraction of SO(2,2).

\section{Conclusions}

\noindent
We reviewed the Goryachev-Chaplygin and Kovalevskaya gyrostats spacetime and the (BENW) pp-wave spacetime. The underlying E(3) group structure for (GCG) and (KG) spacetimes is presented and used to determine higher order St\"{a}ckel-Killing tensors as Eisenhart lightlike lifts of the constants of motion of the classical integrable systems. The corresponding quantum cases are considered. In general the wave operator must be augmented by quantum corrections which are not always expressible in purely geometric terms. It is pointed out the resemblance of the gyrostat cases to the corresponding top cases, where the hamiltonians differ only by a few linear terms in the body frame. 

\medskip
\noindent
The Brdi\v{c}ka-Eardley-Nappi-Witten spacetime is governed by the Cangemi-Jackiw group (the metric is bi-invariant on this group) as it was shown elsewhere and displays a number of Killing vectors and higher rank St\"{a}ckel-Killing tensors, which were determined using an algorithm similar to the van Holten algorithm in[9]. It is pointed out that the spacetime of the associated St\"{a}ckel-Killing group is the Newton-Hooke group.

\bigskip
\noindent
{\Large{\bf Acknowledgements}}

\medskip
\noindent
C.R. acknowledges the support of a POS-DRU European fellowship and thanks DAMTP for their kind hospitality, while this paper was written.

\bigskip
\noindent
{\it Bibliography}

\medskip
\noindent
1. {\it  Kubiz$\check{n}\acute{a}$k, D.} Hidden symmetries of higher-dimensional rotating black holes, PhD thesis, 2008 arXiv 0809.2452

\medskip
\noindent
2. {\it Eisenhart, L. P.} Dynamical trajectories and geodesics, {\bf Annals Math. 30 (1928) 591}

\medskip
\noindent
3. {\it Komarov, I. V.} Goryachev-Chaplygin Top in Quantum Mechanics, {\bf Teoreticheskaya i Matematicheskaya Fizika, Vol. 50, No. 3, pp. 402-409, March 1982}

\medskip
\noindent
4.{\it Komarov, I. V.} Exact solution of the Goryachev-Chaplygin problem in quantum mechanics, {\bf J. Phys. A: Math. Gen. 15 (1982) 1765-1773}

\medskip
\noindent
5.{\it Borisov, A. V., Mamaev, I. S., Kholmskaya, A. G.} Kovalevskaya Top and Generalizations of Integrable Systems, {\bf Regular and Chaotic Dynamics, 2001 Volume 6 Number 1}, arXiv:nlin/0504002

\medskip
\noindent
6.{\it Bolsinov, A. V., Fomenko, A. T.} Integrable geodesic flows on the sphere, generated by Goryachev-Chaplygin and Kowalewski systems in the dynamics of a rigid body, {\bf Mathematical Notes, vol. 56, nos. 1-2, 1994}

\medskip
\noindent
7.{\it Gibbons, G. W., Houri, T.,  Kubiz$\check{n}\acute{a}$k, D., Warnick, C.M.} "Some Spacetimes with Higher Rank Killing-St\"{a}ckel Tensors", arXiv:1103.5366

\medskip
\noindent
8. {\it Kruglikov, B.} Invariant characterization of Liouville metrics and polynomial integrals,{\bf J.geomphys vol. 58 issue 8 2008 979-995} arXiv:0709.0423

\medskip
\noindent
9. {\it van Holten, J. W.} Covariant hamiltonian dynamics {\bf Phys Rev. D 025027 (2007)} hep-th/0612216

\medskip
\noindent
10.{\it Vi\c{s}inescu, M.} Higher order first integrals of motion in a gauge covariant Hamiltonian framework, {\bf Mod.Phys.Lett. A25:341-350,2010} arXiv:0910.3474 

\medskip
\noindent
11.{\it Gibbons, G. W., Pope, C. N.} Kohn's Theorem, Larmor's Equivalence Principle and the Newton-Hooke Group, {\bf Annals Phys.326:1760-1774,2011} arXiv:1010.2455

\medskip
\noindent
12.{\it Komarov, I. V., Zalipaev, V. V.} The Goryachev-Chaplygin gyrostat in quantum mechanics, {\bf J. Phys. A: Math. Gen. 17 (1984) 1479-1488}

\medskip
\noindent
13. {\it Komarov, I. V.} A Generalization of the Kovalevskaya Top, {\bf Phys. Lett. A Vol. 123 no. 1 (1987) 14-15}

\medskip
\noindent
14.{\it Gibbons, G. W., Patricot, C. E.} Newton-Hooke spacetimes, Hppwaves and the cosmological constant, {\bf Class.Quant.Grav.20:5225,2003}   hep-th/0308200

\medskip
\noindent
15.{\it Gibbons, G. W., Herdeiro, C. A. R.} A simple non-commutative curved spacetime, 2001, (unpublished)

\medskip
\noindent
16.{\it Wolf, K. B., Vicent, L. E.} The Fourier U(2) Group and Separation of Discrete Variables, {\bf SIGMA 7 (2011), 053}

\medskip
\noindent
17. {\it Chanu, C., Degiovanni, L., McLenaghan, R. G.} Geometrical classification of Killing tensors on bidimensional flat manifolds,{\bf J. Math. Phys. 47, 073506 (2006)}, arXiv:math/0512324

\end{document}